\documentstyle[epsfig,aps]{revtex}
\newcommand{\bba}{\begin{eqnarray}}
\newcommand{\eea}{\end{eqnarray}}
\newcommand{\bb}{\begin{equation}}
\newcommand{\ee}{\end{equation}}
\newcommand{\bban}{\begin{eqnarray*}}
\newcommand{\eean}{\end{eqnarray*}}

\def\a{\alpha}
\def\b{\beta}

\def\d{\delta}

\def\f{\phi}

\def\z{\psi}

\def\m{\mu}

\def\p{\pi}

\def\r{\rho}
\def\s{\sigma}

\def\t{\tau}

\def\G{\Gamma}

\def\S{\Sigma}

\date{8 December 2000}
\tighten

\begin{document}
\title{On Cosmological Perturbations on a Brane in an Anti--de Sitter Bulk}
\author{C. van de Bruck${}^1$\thanks{Electronic address: C.VanDeBruck\,@\,damtp.cam.a
c.uk} and 
M. Dorca${}^2$\thanks{Electronic address: dorca\,@\,het.brown.edu}}
\address{${}^1$ Department of Applied Mathematics and Theoretical Physics, 
Center for Mathematical Sciences,\\
University of Cambridge, Wilberforce Road, Cambridge CB3 0WA, U.K.}
\address{${}^2$ Department of Physics, Box 1843, Brown University, Providence, RI 029
12, USA}
\noindent
\maketitle
\begin{abstract}
In this paper, we consider
cosmological perturbations on a brane universe embedded in 
an Anti--de Sitter bulk. We use a novel gauge, in which the 
full five--dimensional problem is in principle solvable. In this
gauge  we derive the equations for scalar, 
vector and tensor perturbations. These equations  are necessary in order to
calculate microwave background anisotropies in this particular
scenario. Throughout the paper, we draw attention 
to the influence of the bulk gravitons, which act as a source for 
the perturbations on the brane. In addition, we find that isocurvature
modes are generated due to the influence of bulk gravitons. 
\end{abstract}
\pacs{Keywords: gravitation, Einstein field equations, membrane
theory, perturbation theory, cosmology}

\section{Introduction}
In a now famous paper, Randall and Sundrum have argued that
compactification is not the only way one should think on
how to hide extra spacetime dimensions \cite{RSII}. In fact, 
rather than  compactifying the 
extra dimension(s), they showed that not only matter could be 
confined on a brane with tension, but also gravity 
itself. In their setup they discussed the case for a brane 
embedded in an Anti--de Sitter spacetime with a $Z_2$--symmetry. 
They found that, due to the curvature of the Anti--de Sitter 
spacetime, the Kaluza--Klein zero mode was indeed confined on 
the brane. Thus, for energy densities lower than the brane tension,
gravity is effectively four--dimensional from the point of view of
an observer confined on the brane. 

The cosmological consequences for the background evolution 
in this type of brane world scenario 
were discussed intensively (for a rather incomplete list see, e.g. \cite{cosbra}). 
However, a definite test for this alternative to compactification is 
still missing. As such, the study of cosmological perturbations in
brane worlds is very important. Indeed, perturbations can evolve
differently in these models. Thus, there is the hope that the 
signs of extra dimensions may  be found in the anisotropies of the 
cosmic microwave background or in the matter distribution of the 
universe (for a discussion on perturbations in brane worlds, see 
\cite{creation}, \cite{ours}, \cite{others} and \cite{ours2}). 

In an interesting geometric approach, Einsteins equations were derived
on the brane \cite{tetsu}. It was found that they can be 
written as follows,

\begin{equation}\label{einsteinbrane}
G_{\mu\nu}^{(4)} = -\Lambda_{{\rm eff}}g_{\mu\nu} + 8 \pi G T_{\mu\nu}
+ \kappa_5^4 \pi_{\mu\nu} - E_{\mu\nu},
\end{equation}
being $\kappa_5$ the five--dimensional gravitational coupling constant.
The terms in this equation have the following physical meaning:

\noindent
i) $g_{\mu\nu}$ is the induced metric on the brane, 

\noindent
ii) $\Lambda_{{\rm eff}}$ is the effective cosmological 
constant on the brane, which has contributions from the bulk 
cosmological constant and brane tension, and could be tuned 
to be zero, 

\noindent
iii) $G$ is Newton's gravitational constant, which 
is directly connected to the brane tension $U_b$ 
by $48 \pi G=\kappa_5^4 U_b$,

\noindent
iv) $\p_{\mu\nu}$ is a tensor quadratic in the stress--energy tensor 
$T_{\mu\nu}$ for the brane matter,

\noindent
v) and $E_{\mu\nu}$ is the projection of the five--dimensional Weyl
tensor onto the brane. This is a non--local contribution which 
physically describes the influence of the bulk gravitons.
For a homogeneous brane universe, 
$E_{\mu\nu}$ gives rise to a ``dark radiation'' term. That is, 
matter with the equation of state $p=\rho/3$. The amplitude of 
this radiation term is arbitrary, and actually it might set to be zero. 
Indeed, it was argued, that this amplitude measures the mass of a 
black hole inside the bulk \cite{bowcock}.

For an homogeneous 
and isotropic brane universe, equation (\ref{einsteinbrane}) can be  
used to derive an effective Friedmann equation, which besides the
usual four--dimensional Friedmann equation, it also contains a term 
quadratic in the energy density on the brane. 

In \cite{Langlois}, the equations for scalar 
cosmological perturbations in such a framework where 
derived. It was shown that the usual four--dimensional equations appear, 
together with a quadratic correction. Again, the influence 
of the bulk appeared in form of a fluid with an equation of state
$p=\rho/3$. However, these equations where derived using gaussian normal 
coordinates on the brane. 
In this paper, we derive the perturbation equations 
in a gauge in which, at least in principle, the full
five--dimensional dynamics is solvable. We find, as expected, that the 
form of Einstein equations is the same as found in \cite{Langlois}. 
In addition to that, we derive the equations for vector
and tensor perturbations, on the believe that they could also be of 
relevance in calculating the anisotropies in the cosmic microwave background.

The paper is organized as follows: after discussing very briefly the 
background equations in the next section, we derive
the perturbed Einstein equations on the brane for scalar
perturbations in section 3. As a result, we explicitly see 
that bulk gravitons act as a source for scalar perturbations. 
Following that, we derive the brane equations for
vector and tensor perturbations in section 4 and section 5. 
We summarize our findings and discuss consequences for structure 
formation in section 6. 

\section{The Background}
In this section, we briefly recapitulate the background equations which 
govern the expansion of the homogenous and isotropic 
brane universe (see \cite{cosbra} and 
\cite{ours2}). We assume that the bulk is Anti--de Sitter, i.e. 
it contains only a negative cosmological constant. 
In particular, we will consider here the case of an isolated brane only, 
i.e. a cosmological version of the second Randall--Sundrum model. 

We use the equations, which where derived in \cite{ours} and \cite{ours2}. 
Restricting the 55--component 
of the background Einstein equations, using the proper time 
$\t$ of the brane (i.e. $d\t =abdt$) and  setting 
$\kappa_5 = 1$, one can easily derive,

\bb\label{fried0}
\frac{a_{\t\t}}{a}+\frac{a_\t^2}{a^2}+\frac{k}{a^2}=-\frac{1}{3}\left[
\frac{1}{12}\r_{b}(\r_{b} +3p_{b})+q\right].
\ee
In this equation, $\rho_p$ and $p_b$ are the total brane energy
density and brane pressure respectively and $q$ is the 55--component of the 
bulk energy--momentum tensor. 
We follow now the Randall--Sundrum construction, in which the 
total brane density is a sum of the matter density plus a 
contribution from the brane tension, i.e. 
\bb
\r_{b}=\r_M+U_B,\;\;\; p_{b}=p_M-U_B,
\ee
where $\r_M$ and $p_M$ denote the energy density 
and pressure of the ``ordinary'' matter on
the brane, respectively, and $U_B$ is the brane tension. 
Then, equation (\ref{fried0}) has the following solution 
(see \cite{cosbra} or \cite{ours2} for details)
\bb
\frac{a_\t^2}{a^2}=\frac{\r_M^2}{36}+\frac{8\p G}{3}\r_M+
\Lambda_{\rm eff}-\frac{k}{a^2}
-\frac{\cal C}{a^4},
\ee
where we identify $U_B=48\p G$, being $G$ the four--dimensional 
Newton's constant. Also,  $\cal C$ is an arbitrary constant and
$\Lambda_{\rm eff}$ is an effective cosmological term given by,
\bb
\Lambda_{\rm eff}=\frac{1}{6}\left(\frac{U_B^2}{6}-q\right).
\ee
Following Randall and Sundrum, we assume that the brane tension 
cancels the contribution from the bulk cosmological constant, 
so that 
\bb
\r = -p = -q =  -\frac{U_B^2}{6}, 
\ee
for which $\Lambda_{\rm eff}=0$. We additionally assume that the 
arbitrary constant ${\cal C}=0$, i.e. we neglect the contribution 
from the projected bulk Weyl tensor. This is justified, given the 
assumptions we make for the symmetries of the bulk \cite{bowcock}.
Thus, the Friedmann equation on the brane that
we will consider is
\bb\label{fried1}
{\cal H}^2=\frac{\r_M^2}{36}+\frac{8\pi G}{3}\r_M
-\frac{k}{a^2},
\ee
where we use the notation ${\cal H}=a_\t/a$ for the Hubble parameter 
on the brane.

\section{The scalar perturbations}

In the following sections we would like to derive 
the effective Einstein equations for first order perturbations
on the brane. In order to do that, we have to restrict the full
Einstein equations onto the brane. In doing so, we will make use of 
the formalism presented in \cite{ours}, where the full
five--dimensional equations can be found. Since we are not going to 
repeat the equations for scalar perturbations here, we refer the interested
reader to \cite{ours} for further details. 

We start with the {\em scalar perturbations}. The perturbed metric we 
use takes the form

\begin{eqnarray}\label{escamet}
ds^2 &=& a^2b^2 \left[(1+2\phi)dt^2 - (1-2\Gamma) dy^2  - 2 W dydt -2Bdydx^i
      \right]-a^2\Omega_{ij}(1 - 2\psi)dx^i dx^j. 
\end{eqnarray}
As explained in \cite{ours}, the brane can be located at 
$y={\rm const.}$ along the fifth dimension in this gauge, and the variable
$B$ contains only information about the anisotropic stress on the
brane. In the case for vanishing anisotropic stress on the brane, the
variable $B$ can be set to be zero everywhere and the metric would 
coincide with the {\it generalized longitudinal gauge}. 

We are going to assume for simplicity that the
bulk cosmological constant and the brane tension are real constants, so that
$\d\r_b = \d\r_{M}$ and $\d p_b= \d p_M$. Also, we will use $\t$ as the
{\em proper time on the brane}, that is

\bb
a(\t)b(\t)=1.
\ee
Then, the  restriction of the perturbed Einstein equations on the brane is 
straightforward but rather tedious. Such a restriction proceeds as follows:

\noindent
i) We first use the background Einstein equations to express the
second $y$--derivatives of the scale factors $a$ and $b$ in terms of
their first $y$--derivatives and first and second $t$--derivatives 
(see \cite{ours} for the background Einstein equations). 

\noindent
ii) Then we substitute the jump conditions for
the first derivatives of $a$, $b$, $\f$, $\z$, $\G$ and the jump
conditions for the functions $W$ and $B$ making sure to take the
lower sign in any $\pm$ or $\mp$ combination (which refers to the
brane on the right hand side, i.e. $y=R$, where we assume the 
brane to be located).

\noindent
iii) Finally, we note that the second $y$--derivatives of 
$\f$, $\z$ and $\G$ together with the
first y--derivatives of $W$ and $B$ are {\em not locally determined} on
the brane. Thus, they measure a non--local interaction of the
brane with the bulk. Following \cite{Langlois},
it is rather convenient to write these terms according to,

\bba
\phi_5 &=& \frac{\left(\phi' a^2b^2\right)'}{a^2b^2},\;\;\;
\psi_5=\frac{\left(\psi' a^2\right)'}{a^2},\\
\Gamma_5 &=& \frac{\left(\Gamma a^2b^2\right)'}{a^2b^2},\;\;\;
W_5=\frac{\left(W a^2b^2\right)'}{a^2b^2},\\
\S_5 &=& \frac{\left(B a^3\right)'}{a^3}.\label{B5}
\eea
Because we are assuming that there are no matter 
perturbations in the bulk, we simply have
$\d G^{\alpha}_{~\beta} = 0$.
Using the junction conditions for the scale factors $a$ and $b$ 
and for the metric perturbations, we find that the scalar part of 
Einsteins equations can be written as

\bba
\d^{\rm (4D)}G^0_{~0} &=& 8 \pi G \d\r_M +\frac{\r_M}{6}\d\r_M+ \d\r_5,\\
\d^{\rm (4D)}G^0_{~i} &=& -\left\{8\pi G a^2 (\rho_M + p_M)v_{|i} 
+ \frac{1}{6}\rho_M a^2 (\rho_M + p_M) v_{|i} + a^2(\rho_M + p_M)v_{5|i} \right\}\\
\d^{\rm (4D)}G^i_{~j} &=& -\left\{\d p_5+\frac{\d\r_M}{6}(\r_M
+p_M)+8 \pi G \d p_M+\frac{\d p_M}{6}\r_M 
\right\} \d^i_{~j} + 8\pi G \sigma^{|i}_{~|j} + \frac{\r_M}{6}
\sigma^{|i}_{~|j} + {{\s_5}^{|i}}_{|j},
\eea
where $\d^{\rm (4D)}G^i_{~j}$ is the four--dimensional Einstein
tensor, obtained with the four dimensional part of the metric
(\ref{escamet}) in which all perturbations are taken to be functions of
the intrinsic brane coordinates $t$ and $x^i$. Also, we
have conveniently defined

\bba
\d\r_5 &=&\frac{\d\r_b}{12}(2\r_b +3p_b)-3\z_5-{\S_5}^{|k}_{~|k}
+\frac{\r_b}{6} {B}^{|k}_{~|k}+3{\cal
  H}W_5+\frac{\r_b}{2}\G_5 +3{\cal H}\G_\t- 
\frac{1}{a^2}\G^{|k}_{~|k} \nonumber \\
&-&\left[3{\cal H}^2+3{\cal
 H}_\t-\frac{1}{12}U_B^2\right]\G,\\
a^2(\rho_M+p_M)v_5 &=& \Gamma_\t - {\cal H}\Gamma + \frac{1}{2}W_{5} +
\frac{a^2}{2}{\partial \S_5\over\partial\t}
+\frac{a^2}{6}\left(2\rho_b 
+ 3p_b \right) B_\t-\frac{\r_b}{12}W+a^2\frac{3}{4}{\cal H}(\r_M +p_M)B\\
\d p_5 &=&\frac{\d p_b}{12}(2\r_b +3p_b)
+2\z_5-\f_5+\frac{2}{3}{\S_5}^{|k}_{~|k}
-\frac{\r_b}{9}{B}^{|k}_{~|k}-
\frac{\partial W_5}{\partial\t}-2{\cal H}W_5+\frac{p_b}{2}\G_5
-\G_{\t\t}-2{\cal H}\G_\t\nonumber\\
 & & -\left[6{\cal H}^2+2{\cal
H}_\t+\frac{4k}{a^2}-\frac{p_b^2}{4}+\frac{U_B^2}{3}\right]\G
+\frac{2}{3a^2}\G^{|k}_{~|k},\\
{\s_5^{|i}}_{|j} &=&
{\S_5^{|i}}_{|j} - \frac{\d^{i}_{~j}}{3}{\S_5^{|k}}_{~|k}
+ \frac{1}{a^2}{\G^{|i}}_{|j} - \frac{\d^i_{~j}}{3a^2}\G^{|k}_{~|k} 
-\frac{\r_b}{6}{\s^{|i}}_{|j}\eea
The four--dimensional Einstein tensor $\d^{\rm (4D)}G^\mu_{~\nu}$ reads
\bba
\d^{\rm (4D)}G^0_{~0} &=& -6{\cal H}^2\f -6{\cal H}\z_\t+
\frac{6k}{a^2}\z+\frac{2}{a^2}\z^{|k}_{~|k},\\
\d^{\rm (4D)}G^0_{~i} &=& \left\{2{\cal H}\f+2\z_\t\right\}_{|i},\\
\d^{\rm (4D)}G^i_{~j} &=&\left\{ -(6{\cal H}^2+4{\cal H}_\t)\f-2{\cal H}\f_\t
-\frac{1}{a^2}\f^{|k}_{~|k}-2\z_{\t\t}-6{\cal H}\z_{\t}
+\frac{1}{a^2}\z^{|k}_{~|k}+\frac{2k}{a^2}\z\right\}\d^i_{~j} 
+\frac{1}{a^2}{\f^{|i}}_{|j}-\frac{1}{a^2}{\z^{|i}}_{|j}.
\eea
In defining the quantities $\d\r_5$, $\d p_5$, $v_5$ and $\sigma_5$,
we have made explicit use of the junction conditions given in
\cite{ours}. Note that these terms contain the information about 
the bulk perturbations. 
The quantities ${\cal H}^2$ and ${\cal H}_\t$ also 
contain information about the brane and bulk matter fields. Indeed,
according to (\ref{fried1}) and ({\ref{fried0}), they can be written as

\bba
{\cal H}^2 &=& \frac{1}{3}\left[\frac{\r_M^2}{12}+\frac{U_B}{6} \right]\r_M
-\frac{k}{a^2},\\
{\cal H}_\t &=&-\left[\frac{\r_M}{12}+\frac{U_B}{12}\right] (\r_M+p_M)
+\frac{k}{a^2}.
\eea

On the other hand the restriction of the $5\mu$--components
of the perturbed Einstein equations on the brane
give the usual energy--momentum conservation for brane perturbations. 
Finally, rather than the 55--component of the perturbed
Einstein equations, we conveniently use the 55--component
of the perturbed Ricci tensor, i.e. $\d R^5_{~5}=0$. This is possible
because the bulk contains only a pure cosmological constant. Thus, 

\bba
\d R^5_{~5} &=& -3\z_5+\f_5-{\S_5}^{|k}_{~|k}+
\frac{\partial W_5}{\partial\t}+3{\cal H}W_5+\frac{1}{6}(\r_b-3p_b)\G_5
+\left[\frac{\d\r_b}{36}-\frac{\d p_b}{12}\right](2\r_b +3p_b)
+\G_{\t\t}+3{\cal
  H}\G_\t-\frac{1}{a^2}\G^{|k}_{~|k}\\
 & &+\frac{\r_b}{6}B^{|k}_{~|k} 
+\left[{\cal H}_\t+5{\cal H}^2+\frac{4k}{a^2}-\frac{p_b^2}{4}
+\frac{13}{36}U_B^2\right]\G=0.
\eea
This equation allows to find the following relationship between
$\d \rho_5$ and $\d p_5$:
\bb\label{eqstate}
\d p_5=\frac{1}{3}\d \r_5.
\ee
Recall that the origin of the terms involving $\d \rho_5$ and $\d p_5$
comes from a non-local interaction between the brane and the bulk
gravitons. Therefore, in view of (\ref{eqstate}), these gravitons
can be seen as a source for the fluctuations on the brane
\cite{Langlois}. In particular, the bulk quantities 
$\Gamma_5$ and $B$ give rise to an induced anisotropic stress on the
brane. 

We remark that, regardless of having a vanishing anisotropic stress
on the brane,
the quantities $\phi$ and $\psi$ no longer commute. This is true due
to the appearance of an effective anisotropic
stress and it is in clear contrast to the usual four--dimensional case.
Indeed, the traceless part of the $ij$--component reads 

\begin{equation}
\phi^{|i}_{~~|j} - \psi^{|i}_{~~|j} = \sigma^{|i}_{5~|j}
+\frac{\r_b}{6}{\sigma^{|i}}_{|j}.
\end{equation}
Assuming that the anisotropic stress on the branes vanishes, we can 
replace $\psi$ in favor of $\phi$ and $\sigma_5$
(observe that ${{\s_5}^{|k}}_{|k}=0$). Then, the 
$00$--component and the trace of the $ij$--component can be written as

\begin{eqnarray}
\frac{\phi^{|k}_{~|k}}{a^2} - 3{\cal H}\phi_{\tau} - 3\left[ {\cal
H}^2 - \frac{k}{a^2} \right] \phi
&=& 4 \pi G \delta \rho_M + \frac{\rho_M}{12}\delta \rho_M 
+\frac{1}{2}\d\rho_5 - 3\left[{\cal H} \sigma_{5\tau} -
\frac{k}{a^2}\sigma_5\right] \\
\phi_{\tau\tau}+ 4{\cal H}\phi_\tau+
\left[3{\cal H}^2 + 2{\cal H}_{\tau} -\frac{k}{a^2}\right] \phi
&=& 4\pi G\delta p_M + \frac{\rho_M}{12}\delta p_M +
\frac{\rho_M+p_M}{12}\delta \rho_M + \frac{1}{2}\d p_5 +
\sigma_{5,\tau\tau} + 3{\cal H}\sigma_{5,\tau} - \frac{k}{a^2}\sigma_5
\end{eqnarray}
Now, decomposing the pressure perturbations as 
\begin{equation}
\delta p_M = c_{s}^2 \delta \rho_M + {\cal T}\delta S, 
\end{equation}
where $\delta S$ is the entropy production and $c_s$ the sound 
velocity, we can combine both equations to get
\begin{eqnarray}
\phi_{\tau\tau} &+& \left[ 4 + 3c_s^2\right]H\phi_{\tau} 
+ \left[ 2{\cal H}_{\tau} + 3H^2\left(1+c_s^2\right) 
-\frac{k}{a^2}\left(1 + 3c_s^2\right)\right]\phi 
- c_s^2\frac{\phi^{|k}_{|k}}{a^2}\nonumber\\
&=& 4\pi G {\cal T} \delta S + \frac{\rho_M}{12}{\cal T}\delta S 
+\frac{\rho_M + p_M}{12}\delta \rho + \frac{1}{2}\left(\d p_5 - c_s^2
\d\rho_5\right) + \sigma_{5,\tau\tau} + 3H(1+c_s^2)\sigma_{5,\tau} 
-\frac{k}{a^2}\left(1 + 3c_s^2\right)\sigma_5. 
\end{eqnarray}
The right hand side corresponds to the usual 4D case, without any 
further corrections. Also, the first term on the left hand side
corresponds to the term usually found in four dimensions, 
which describes entropy production. Note, that the combination 
$\delta \rho_5 - c_s^2 \delta p_5$ describes an
effective entropy production, induced by bulk gravitons. 
In addition, we find the usual corrections quadratic in the 
energy density and terms involving the effective 
anisotropic stress induced from bulk gravitons. Since we cannot a priori neglect 
these extra terms, which contain information from the bulk, 
a study of the full set of five--dimensional equations is 
required.

\section{The Vector Perturbations}
In this section we draw our attention to the vector
perturbations. It is commonly assumed that since these type of
perturbations decay due to the cosmological expansion, they are
not relevant for the study of cosmological  perturbations.
However, this may not be true in the context of brane world
models. The reason is that, analogously to the case of scalar perturbations,
vector and tensor perturbations are also sourced by bulk gravitons.
Thus the study of both vector and tensor perturbations may be also necessary. 

The most general metric for vector perturbations in brane
worlds with a brane located at a constant $y$ position along the
bulk is,

\begin{equation}\label{permetricvec0}
ds^2 =  a^2\left\{\gamma_{ab} dy^a dy^b 
       -\left[ \Omega_{ij}+
       2F_{(i|j)}\right] dx^i dx^j-2 S_{ai}dy^a dx^i\right\},
\end{equation}
where the three--vectors $S_{ai}$ and $F_i$ have vanishing divergences,
i.e. 
\bb
{{S_a}^i}_{|i}=0\;\;\; ,{{F}^i}_{|i}=0.
\ee

In five dimensions, there exist two gauge--invariant vector
perturbations (see e.g. \cite{ours}). In the generalized longitudinal
gauge, these two gauge--invariant vector perturbations can be
identified to $S_{ai}$, while the remaining vector perturbation $F_i$
can be conveniently cancelled. Thus, working in this gauge, we have

\begin{equation}\label{permetricvec1}
ds^2 =  a^2\left\{b^2 dt^2 -b^2 dy^2 
       -\Omega_{ij}dx^i dx^j-2 S_{0i}dt dx^i-2 S_{5i}dy dx^i\right\}\; ,
\end{equation}
On the other hand, the perturbed stress-energy tensor
for vector perturbations on the brane is given by 
\bb
 \d {T^{\a}}_\b = \left(\begin{array}{ccc}
       0&-(\r_{b}+p_{b})b^{-2}V_{j}&0\\
-(\r_{b}+p_{b})S_0^{i}+(\r_{b}+p_{b})V^{i}&2{\cal F}^{(i}_{~|j)}&U^{i}\\
0&p_{b}b^{-2}S_{5j}-b^{-2}U_{j}&0\end{array}\right)\; ,
 \label{dTVn}
\ee
where $V^{i}$ and $U^{i}$ describe two  ``velocity'' fields for
the matter on the branes.
In the bulk, the first order Einstein equations for the vector 
perturbations are

\bba
a^2b^2\d G^0_{~i} &=& -\frac{1}{2b^2}\left[
\frac{\partial^2}{\partial y^2} +
\left(3\frac{a'}{a}-2\frac{b'}{b}\right)\frac{\partial}{\partial y}
+4b^2k\right] S_{0i}-\frac{1}{2}{S_{0i}}^{|k}_{~|k}+
\frac{1}{2b^2}\left[
\frac{\partial^2}{\partial y\partial t} +
\left(3\frac{a'}{a}-2\frac{b'}{b}\right)\frac{\partial}{\partial t}
\right] S_{5i}=0,\\
a^2b^2\d G^i_{~j} &=& \left[\frac{\partial}{\partial t}+
3\frac{\dot a}{a}\right] {S_0^{(i}}_{|j)}-
\left[\frac{\partial}{\partial y}+
3\frac{a'}{a}\right] {S_5^{(i}}_{|j)}=0\\
a^2b^2\d G^5_{~i} &=& \frac{1}{2b^2}\left[
\frac{\partial^2}{\partial y\partial t} +
\left(3\frac{\dot a}{a}-2\frac{\dot b}{b}\right)\frac{\partial}{\partial y}
\right] S_{0i}
-\frac{1}{2b^2}\left[
\frac{\partial^2}{\partial t^2} +
\left(3\frac{\dot a}{a}-2\frac{\dot b}{b}\right)\frac{\partial}{\partial t}
-4b^2k\right] S_{5i}+\frac{1}{2}{S_{5i}}^{|k}_{~|k}=0.
\eea

The junction conditions can easily found to be
\bba
S_{5i} &=& {\cal F}_i,\label{jumpV1}\\
S_{0i}' &=& {\dot S}_{5i} +
(\r_{b}+p_{b})V_i,\label{jumpV2}\\
{\cal F}_i &=& \frac{U_i}{p_b}.\label{jumpV3}
\eea
Observe, thus, that the velocity field $U_i$ and the matter
perturbation ${\cal F}_i$ are not independent quantities but they
are directly related by (\ref{jumpV3}).
Defining
 
\begin{eqnarray}
{\cal P}_{5i} &=& \frac{(S_{5i}a^2)'}{a^2},\\
{\cal P}_{0i} &=& \frac{1}{2a^5}\left[\frac{a^5}{b^2}\left(
S_{0i}'-{\dot S}_{5i}\right) \right]' \;,
\end{eqnarray}
the first order Einstein equations for vector perturbations on
the brane can be written as
\bba
\d^{\rm (4D)}G^0_{~i} &=& -8\pi G a^2 (\rho_M + p_M)V_i 
- \frac{\rho_M a^2}{6}(\rho_M+p_M)V_i - {\cal P}_{0i} \\
\d^{\rm (4D)}G^i_{~j} &=& 8\pi G\,{{\cal
F}^{(i}}_{|j)}+\frac{\r_M}{6}{{\cal F}^{(i}}_{|j)}+{\cal P}_5^{(i}}_{|j),
\eea
where again, $\d^{\rm (4D)}G^\mu_{~\nu}$ denote the four--dimensional Einstein
tensor, which reads
\bba
\d^{\rm (4D)}G^0_{~i} &=& -\frac{1}{2}{S_{0i}}^{|k}_{~|k}-2kS_{0i},\\
\d^{\rm (4D)}G^i_{~j} &=& \left[\frac{\partial}{\partial\t}+3{\cal
    H}\right]{S_0^{(i}}_{|j)}.
\eea
Finally, the remaining non-zero component of Einstein equations,
$\d G^5_{~i}$, gives the following energy conservation equation for
vector perturbations on the brane,
\bb
(\rho_M + p_M)\dot{V}_i + 
\left[ 2H(\rho_M + p_M) + \dot{p}_M\right]V_i 
+ \frac{{\cal F}^{|k}_{i~|k}}{a^2} + \frac{4k}{a^2}{\cal
F}_i= 0. 
\ee

\section{The Tensor Perturbations}

Tensor perturbations may be physically related
to linearized gravitational waves. For the case of a single extra
dimension the most general metric is 

\begin{equation}\label{permetricten0}
ds^2 =  a^2\left\{b^2 dt^2 -b^2 dy^2 
       -\left[ \Omega_{ij}+
     h_{ij}\right] dx^i dx^j\right\}\; ,
\end{equation}
The 3--dimensional tensor $h_{ij}$, which is already gauge--invariant, is
traceless and divergenceless, i.e.
 
\bb
{{h}^k}_{k}=0\;\;\;\; {\rm and} \;\;\;{{h}^{ij}}_{|i}=0.
\ee
We use now the following perturbed stress-energy tensor
for tensor perturbations on the brane,

\bb
 \d {T^\a}_\b = \left(\begin{array}{ccc}0 &0& 0\\
         0&{\Pi^{i}}_{j}&0\\
              0&0&
              0\end{array}\right)
 \label{dTHn}
\ee
where $\Pi^i_{~j}$ is a 3--dimensional tensor describing the
tensorial part of the anisotropic stress of the matter fields 
on the brane.

The first order Einstein equations for the tensor perturbations 
in the bulk take the very simple form

\bb
\d G^i_{~j}=-\frac{1}{2}\left[\Box -\frac{2k}{a^2} \right] h^i_{~j}=0,
\ee
where $\Box$ stands for the 5--dimensional Laplacian, given by
\bb
\Box =\nabla^\m\nabla_\m =\frac{1}{a^2b^2}\left[
\frac{\partial^2}{\partial t^2}-\frac{\partial^2}{\partial y^2}
+3\frac{\dot a}{a}\frac{\partial}{\partial t}
-3\frac{a'}{a}\frac{\partial}{\partial y}-b^2\partial^k\partial_k
\right].
\ee
Then the junction conditions read
\bb\label{jumpH}
h_{ij}'=\Pi_{ij}.
\ee
If we now define the quantity
\bb
{\cal I}_{ij} =- \frac{\left(h_{ij}' a^5\right)'}{2a^5}, 
\ee
which describes the influence on the brane of the bulk tensor perturbations, 
we can write the effective four--dimensional Einstein equation as
\bb
\d^{\rm (4D)}G^0_{~i} = 8\pi G \Pi^i_{~j} + \frac{\rho_m}{6}
\Pi^i_{~j} + {\cal I}^i_{~j}.
\ee
Once more, $\d^{\rm (4D)}G^\mu_{~\nu}$ denote the four--dimensional Einstein
tensor, which reads
\bb
\d^{\rm (4D)}G^i_{~j}=-\frac{1}{2}\left[\Box_4 -\frac{2k}{a^2} \right]
h^i_{~j}\; ,  
\ee
where
\bb
\Box_4 =\left[
\frac{\partial^2}{\partial \t^2}
+3{\cal H}\frac{\partial}{\partial \t}
-\frac{1}{a^2}\partial^k\partial_k
\right].
\ee

\section{Consequences for Structure Formation}

In this paper we have derived the  equations for scalar,
vector and tensor perturbations on a brane universe embedded in an 
Anti--de Sitter space. We have assumed a single brane and a bulk
without matter perturbations. However, the equations are 
written in such a way, that, at least in principle, the full set of 
5D equations could be solved. As such, the results could be extended 
to include a second brane, for instance. 

What was found is that all kind of perturbations are sourced by 
bulk gravitons. This is not surprising, since perturbations on the
brane induce perturbations in the bulk and vice versa. However,
for structure formation on the brane, the dynamics of these source
terms should be understood. In the particular case of scalar
perturbations, the non--local interaction between the brane and the
bulk is described by an effective energy density, pressure, velocity 
field and anisotropic stress. 
The effective pressure and energy density are related by an 
{\em radiation like} equation of state. Furthermore, the combination 
$\delta p_5 - c^2_s \delta \rho_5$, where $c_s$ is the {\em matter 
sound speed}, describes an effective entropy production. This 
entropy production can in principle generate isocurvature modes. 
The efficiency of the production depends strongly on the evolution of 
$\delta \rho_5$, $\delta p_5$ and the sound speed $c_s$. 
On the other hand, within the formalism presented so far, 
the evolution of $\sigma_5$ is not constrained. 
Although the contribution from bulk gravitons (i.e. the
dark radiation term) may be neglected for the background, it
reappears in the cosmological perturbation equations. 
There, we cannot arbitrarily set its amplitude to zero. 
One important consequence is that, contrary to 
the usual four--dimensional case, the metric variables $\phi$ and 
$\psi$ do not necessarily commute, even in the absence of anisotropic
stress on the brane. Thus, given that the stress history is very important 
for structure formation (see \cite{HU}), the evolution of $\sigma_5$
has to be studied carefully. 

Besides scalar perturbations, we have also found the equations for
vector and tensor perturbations. As expected, these perturbations 
are sourced by bulk gravitons as well. This means that any 
constraints given to this particular model (from the CMB, for example) 
have to take into account the contribution of tensor and, once
produced, vector modes also. Of course, in order to have the 
information about bulk perturbations, 
one has to solve the full Einstein equation in the bulk. 
Although the gauge used here makes the 5D equations 
in principle solvable, this is a task beyond 
the scope of the present work. 

In \cite{Langlois} the connection to ``seed''--models of
structure formation was made\footnote{That the model has similarities
with defect models of structure formation was independently pointed
out by Richard Battye to us.}. However, within the model we discuss, 
it is not clear, if the seeds obey a scaling solution. Non--scaling 
seeds are rather peculiar and need to be studied further (for the case
of non--scaling cosmic strings, see \cite{strings}). In addition,
the projection onto the brane is non--local, which is another new effect in
this kind of brane world scenarios. Thus, at this stage, it is not at 
all clear if the model could be ruled out by current observations. 

Finally we would like to comment what would change if a second brane
(with negative tension) is somewhere else in the bulk. In that
case, all the quantities describing the influence of bulk gravitons 
(i.e. $\d \r_5$, $\d p_5$,
etc.) would carry the information about the perturbations on the other 
brane. Indeed, through the full 5D equations, it would be possible 
to relate $\d \r_5$, $\d p_5$, etc. to the matter perturbations on the 
other brane. However, due to the warp factor these might be strongly
suppressed if we lived on the positive tension brane. If that is the 
case, the low--effective theory relating perturbations of the matter
fields on both branes would be, when the physical distance of the 
branes is stabilized, the usual four--dimensional Einstein gravity, 
with contributions from both branes. 

\vspace{1cm}

\noindent{\it Acknowledgments}: We are grateful to Richard Battye, 
Robert Brandenberger, Andrew Mennim and Toby Wiseman for useful 
discussions. C.v.d.B. was supported by the Deutsche
Forschungsgemeinschaft (DFG). M. Dorca is supported by
the {\em Fundaci\'on Ram\'on Areces}. The research was supported
in part (at Brown) by the U.S. Department of Energy under Contract 
DE-FG02-91ER40688, TASK A.

\end{document}